\documentclass[prl,twocolumn,showpacs,amsmath,amssymb,superscriptaddress]{revtex4}
\usepackage{pifont,amssymb,epsf,graphicx}
\begin{document}
\title{Reply to Comment on "Nonmagnetic Impurity Resonances as a Signature of Sign-Reversal Pairing
in FeAs-based Superconductors" }
\author{Degang Zhang}
\affiliation{Texas Center for Superconductivity and Department of
Physics, University of Houston, Houston, TX 77204, USA}


\maketitle

In Ref. [1], starting from two Fe ions per unit cell and two
degenerate orbitals per Fe ion, I have constructed a two-orbital
four-band tight-binding model, which describes correctly the
characteristics of the Fermi surfaces in the FeAs-based
superconductors. Most recently, based on this model, we have
investigated antiferromagnetism and superconductivity in
electron-doped samples by employing Bogoliubov-de Gennes equation
[2]. It is shown that the coherence peak at positive bias energy is
lower than that at negative energy for over-doped samples while the
coherence peak at positive energy becomes higher due to the
coexistence of spin density wave and superconductivity for
under-doped samples. Meanwhile, the heights of the coherence peaks
at negative and positive bias energies are approximately equal for
optimal-doped samples. The results are consistent with recent
scanning tunneling microscopy (STM) observations [3]. To the best of
my knowledge, no other models can fit all the features of local
density of states (LDOS) up to now.

In the preceding Comment [4], Daghofer and Moreo pointed out that
the energy band structure in Ref. [1] is not degenerate at $\Gamma$
point, which is contradictory to local density approximation or a
Slater-Koster approach. Here I emphasize that the two-orbital
four-band tight-binding model in Ref. [1] is built to fit angle
resolved photoemission spectroscopy (ARPES) experiments rather than
other theoretical calculations. The authors of this Comment seem to
miss the important ARPES observations in over-doped samples, where
the $\alpha$ band disappears, but the $\beta$ band still exists (see
Fig. 3 in Ref. [5] and Figs. 3 and 4 in Ref. [6]). Therefore, it is
obvious that the local density approximation or Slater-Koster
approach is inconsistent with the experimental facts near $\Gamma$
point. The disappearance of $\alpha$ band leads to the asymmetric
coherence peaks in LDOS [2], which were also observed by the STM
experiments [3,7,8].

The authors of this Comment also argued that $t_2$ and $t_3$ in Ref.
[1] are equal due to the symmetric hopping paths. However, this is
only their opinion. It is known that the FeAs-based superconductors
are very complex materials due to electron or hole doping, magnetic
moment on Fe ion, and lattice distortions. In inelastic neutron
scattering experiments [9, 10], the magnetic excitation structure
exhibits strong $C_2$ $(180^\circ)$ symmetry such that the nearest
neighbor exchange constants along $a$ and $b$ axes are dramatically
different due to tiny lattice distortions. In Ref. [11], Nascimento
{\it et al.} found different Fe-As orbital structure associated with
distinct As ions. In recent STM experiments [12], Chuang {\it et
al.} have also observed the local electron states with $C_2$ rather
than $C_4$ $(90^\circ)$ symmetry. This local electron property is
not produced merely due to crystal symmetry [12]. The inequality of
$t_2$ and $t_3$ could reflect these experimental facts. The
difference of $t_2$ and $t_3$ determines the sizes of hole pocket
around the $\Gamma$ point and electron pocket around the ${\rm M}$
point, depending on the samples studied.

The author would like to thank C. S. Ting, S. H. Pan, Ang Li, and
Tao Zhou for useful discussions, and especially S. H. Pan and Ang Li
for providing me their STM data. This work was supported by the
Texas Center for Superconductivity at the University of Houston and
by the Robert A. Welch Foundation under Grant No. E-1146.


\end{document}